\documentstyle[psfig,editedvolume]{crckapb}

\def\ros{{\sl ROSAT }}

\def\approxlt{\mathrel{\hbox{\rlap{\lower.55ex \hbox {$\sim$}}
        \kern-.3em \raise.4ex \hbox{$<$}}}}
\def\approxgt{\mathrel{\hbox{\rlap{\lower.55ex \hbox {$\sim$}}
        \kern-.3em \raise.4ex \hbox{$>$}}}}

\begin{opening}
\title{X-ray emission mechanisms of LINERs}  
\author{Stefanie Komossa}
\institute{}
\author{Dieter Breitschwerdt} 
\institute{}
\author{Hans B\"ohringer}
\institute{Max-Planck-Inst. f\"ur extraterrestrische Physik, Postfach 1603,
           D-85740 Garching, Germany;~~ skomossa@xray.mpe.mpg.de}
\author{Janek Meerschweinchen}
\institute{Weststrasse 19, 3063 Obernkirchen\,2, Germany}

\end{opening}

\runningtitle{X-ray emission mechanisms of LINERs}

\begin{document}

% The \begin{document} command comes after the \end{opening}
% command.

\vspace*{-10cm}
\begin{verbatim}
Contribution to the proceedings of `Astrophysical Dynamics'
(Evora, April 14-16, 1999); to appear in Ap&SS.
Preprint available at http://www.xray.mpe.mpg.de/~skomossa
\end{verbatim}
\vspace*{8.1cm}

\begin{abstract}
We present here an analysis of the X-ray properties of a sample of
LINER galaxies observed with the {\sl ROSAT} PSPC and HRI instruments.
A spatial analysis shows that the bulk of the X-ray emission is
consistent with arising from a point source; some extended emission
appears at weak emission levels.
The X-ray spectra are formally best described by a powerlaw with photon index
$\Gamma_{\rm x} \approx -2$ or thermal emission from a Raymond-Smith plasma
with highly subsolar abundances ($Z \leq 0.1$).
Several emission mechanisms that might contribute to the observed X-ray spectra
are discussed. In particular, we take the very subsolar abundances derived
from Raymond-Smith fits as an indication of a more complex emission mechanism,
like the presence of a second hard component or plasma out of equilibrium.
\end{abstract}

\section{Introduction}

Low-ionization nuclear emission line regions, LINERs,
are characterized by their optical emission line spectrum
which shows a lower degree of ionization
than active galactic nuclei (AGN).
Their major power source and line excitation mechanism
have been a subject of lively debate ever since their discovery.
If powered by accretion, they probably represent the low-luminosity end
of the quasar phenomenon
and their presence has relevance to, e.g., the evolution of quasars,
the faint end of the Seyfert luminosity function, the soft X-ray background,
and the presence of SMBHs in nearby galaxies. A detailed study
of the LINER phenomenon is thus very important.
X-rays are a powerful tool to investigate the presence of an
active nucleus within the LINER galaxies via X-ray variability
and luminosity,  and to explore the physical properties
of LINERs in general.
An analysis of {\sl ASCA} spectra of several low-luminosity
AGN including 5 LINERs (Ptak et al. 1999) revealed the presence of a hard
powerlaw--like emission component plus soft thermal emission.

Here, we present a study of 13 LINERs observed in the
soft X-ray spectral region (0.1--2.4 keV) with the instruments on-board
{\sl ROSAT}. All of them were targets during the \ros all-sky survey,
some of them detected for the first time in X-rays, 
and for 8 of them additional pointed PSPC and/or HRI observations are available.
Luminosities given below are based on $H_{0}$=75\,km/s/Mpc; distances of
very nearby galaxies were taken from Tully's (1988) catalog.

%-----------------------------------------------------------------
\begin{figure*}
\hspace*{-0.5cm}
     \vbox{\psfig{figure=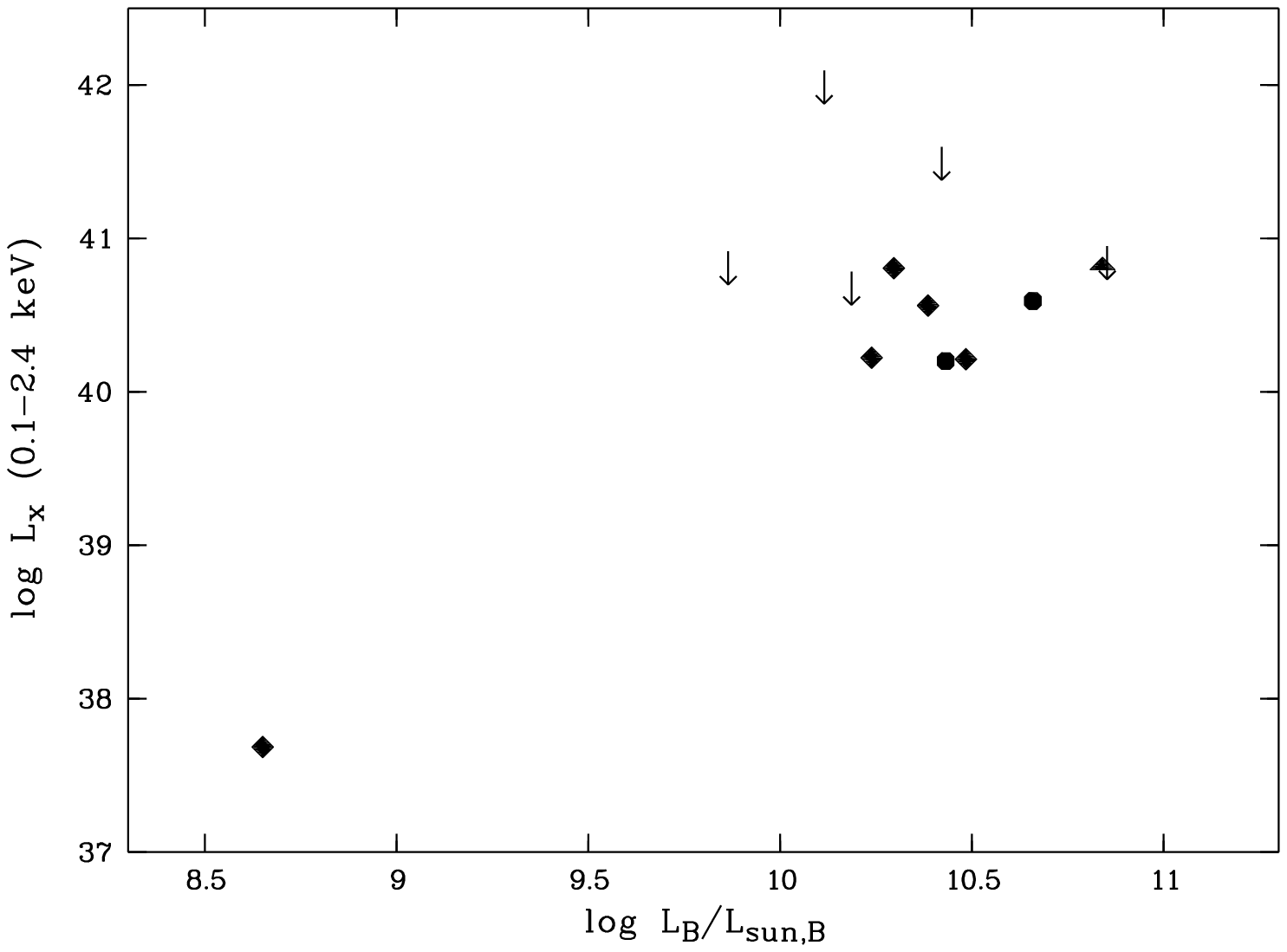,width=7.7cm,%
          bbllx=1.9cm,bblly=1.1cm,bburx=18.3cm,bbury=12.2cm,clip=}}\par
     \vspace*{-5.0cm}\hspace*{7.38cm}
   \psfig{file=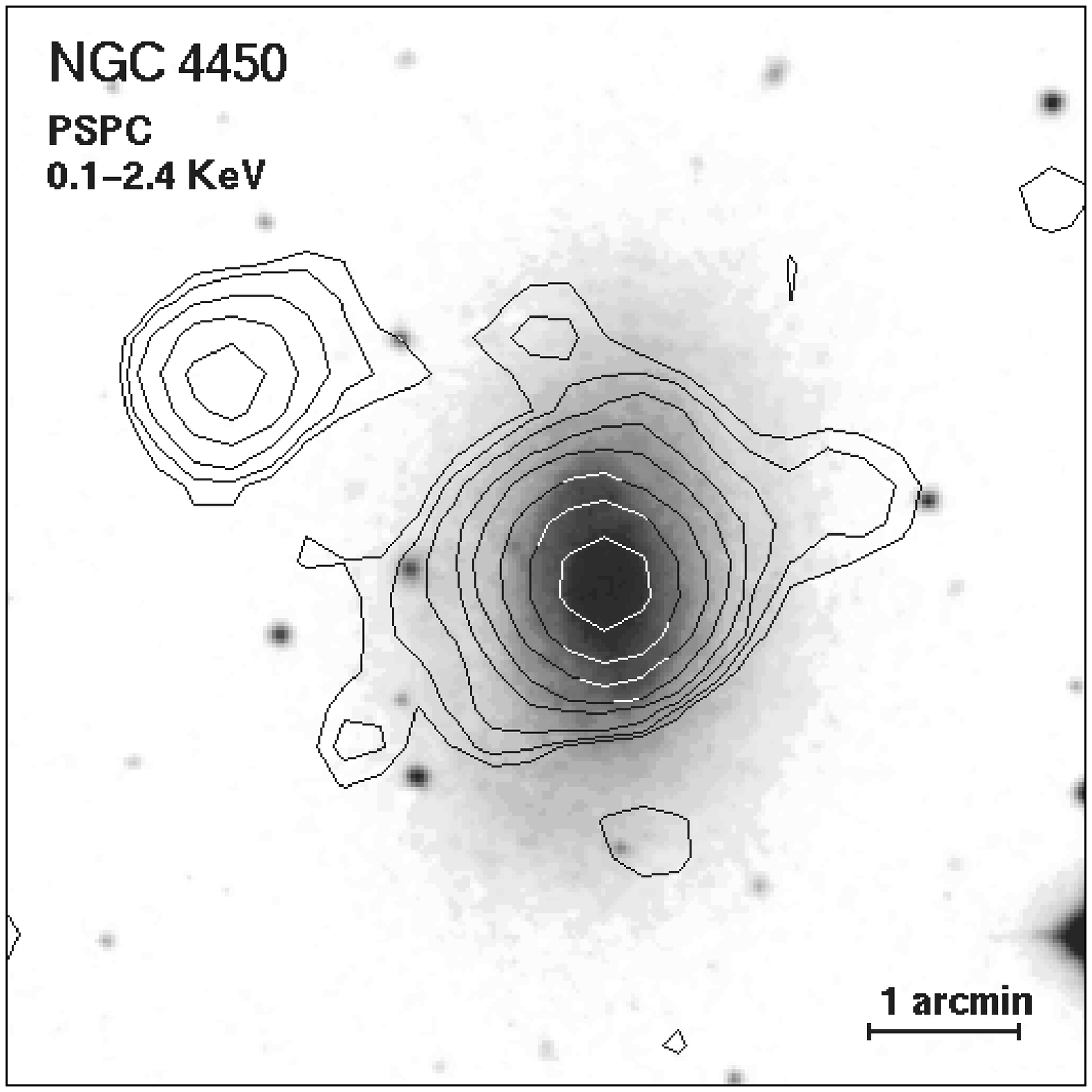,width=4.4cm}
\hspace*{0.6cm}
  \caption[]{ {\em Left}: LINERs in the $L_{\rm x} - L_{\rm Blue}$ diagram.
The filled symbols denote morphological galaxy types.
Circles: S0s (NGC\,4036, NGC\,2768);
lozenges: SAs (NGC\,404, NGC\,3898, NGC\,4450, NGC\,6500, NGC\,3642);
triangle: SAB (NGC\,5371); arrows
mark upper limits (NGC\,4419, NGC\,5851, NGC\,5675, IC\,1481, NGC\,1167);
all galaxy identifications are given from left to right. 
{\em Right}: Contour plots of the X-ray emission of the LINER NGC\,4450
overlaid on an optical image from the digitized POSS.
}
\label{lxlb}
\end{figure*}
%----------------------------------------------------

\section{Results of the X-ray analysis}

The results can be summarized as follows 
(see Komossa et al. 1999 for the full analysis):

\begin{itemize}
\item X-ray luminosities  
 range between $\log L_{\rm x} = 37.7$ (NGC\,404)
and 40.8 (NGC\,4450). The ratios $L_{\rm x}$/$L_{\rm B}$, when compared to
early-type galaxies, are located in the intermediate region of
the $L_{\rm x}-L_{\rm B}$ diagram (Fig. 1; see Komossa et al. 1999
for an explicit comparison with elliptical galaxies)
and similar emission mechanisms may contribute to the observed X-ray 
luminosities.

\item Whereas the bulk of the X-ray emission from the LINERs
is consistent with arising from a point source 
there is some extended emission seen at weak emission levels in some sources.

\item We do not detect any short timescale (hours to weeks)
temporal variability.
This is consistent with the earlier suggestion that LINERs may
accrete in the advection-dominated mode.
Only one source, NGC\,2768, is slightly variable on a timescale of months.

\item The X-ray spectra of the LINER galaxies show some spectral
variety. Half of them are best described by a powerlaw of photon index
$\Gamma_{\rm x} \simeq -2$
whereas the other ones are best matched by a Raymond-Smith model with $kT 
\simeq$ 0.6 keV
and heavily depleted metal abundances.
Below, we discuss possible explanations for the unusually low inferred 
abundances. 

\end{itemize}

\section{X-ray emission mechanisms}

Some sources are best described by a single AGN-like powerlaw
with X-ray luminosity above that expected from discrete stellar
sources. These spectra most likely indicate the presence of
low-luminosity AGNs in the centers of these LINER galaxies.
We concentrate here on possible explanations for the very low metal abundances
(1/10 -- 1/20 $\times$ solar) formally derived for the other sources
when a Raymond-Smith model is fit. 

{\em \underline{Presence of a second hard component.}}~~
One way to avoid heavily depleted abundances is to invoke a second hard X-ray 
spectral
component (see the discussion in Buote \& Fabian 1998, Komossa \& Schulz 1998, 
and
Komossa et al. 1998 for details).
The second component could be of powerlaw shape and originate from
a low-luminosity AGN (LLAGN),     
or it could be hot thermal emission from an extended component. The two 
possibilities
cannot be distinguished by spectral fits to \ros data.  

Alternatively, the emitting gas might show a range of temperatures.  
Strickland \& Stevens (1998) carefully discussed this possibility
by simulating an expanding superbubble under the assumptions of axial 
symmetry and collisional ionization equilibrium (CIE). 
They find that by fitting a single-temperature model to 
\ros data instead of a multi-temperature one, the true metal abundances
are heavily underestimated.  

However, nature may be even more complicated. If the plasma is not in 
CIE, the density and temperature structure of the emitting region 
cannot be reproduced by a finite one-parameter family of curves with 
merely differing temperatures, since each ionization stage of each ion 
would have its `own temperature'. 

{\em \underline{Non-equilibrium effects in a hot extended plasma component.}}~~
When applying Raymond-Smith or similar codes, the assumption of CIE
has to be kept in mind. 
In optically thin plasmas such an assumption can never be strictly fulfilled, 
since collisional ionization and radiative recombination are not 
statistically {\em inverse processes}, and therefore can only hold in an 
approximate sense. If in addition the plasma 
is in motion, e.g. due to its origin in a superbubble or galactic wind, 
the adiabatic cooling timescale can be shorter than any timescale associated 
with atomic processes. The resulting spectrum may then be dominated by 
a strong contribution from {\em delayed recombination} of highly ionized 
species (Breitschwerdt \& Schmutzler 1994, 1999). The kinetic temperature 
of the electrons in the X-ray emitting plasma is typically much lower than 
in a CIE plasma, and collisional ionization contributes 
mainly to very soft X-ray or UV emission. Thus a two- or multitemperature 
plasma can be mimicked, as far as the spectral resolution of current X-ray 
detectors is concerned. 
Sub-solar `Raymond-Smith abundances' can also be obtained with
non-CIE spectra from a galactic outflow, folded through the
ROSAT detector response matrix,
as we have verified by preliminary calculations of non-CIE models. 

In conclusion, both effects (the presence of a LLAGN and non-equilibrium plasma)
are likely to play a role to some extent.
The importance of the latter also depends on the origin of the gas. 
The galaxies of the present sample are mostly type S0 and SA. Whereas
in ellipticals the extended gas is `old' and the standard equilibrium assumptions
may be justified, in spirals a contribution from a young starburst
that drives superwinds is well possible. In that case the gas is likely 
to be far from equilibrium 
(see the discussion in Breitschwerdt et al. 1999 and these proceedings). 

Given the spectral variety of LINERs with contributions from several
emission components, future studies of both, spectra of individual objects
as well as larger samples will certainly give further insight into
the LINER phenomenon which provides an 
important link between active and `normal' galaxies.

% \begin{acknowledgements}
% St.K. acknowledges support from the Verbundforschung under grant No.
% 50\,OR\,93065.
% \end{acknowledgements}

{}

\end{document}